\documentclass[12pt]{article}

\usepackage{amsmath,amssymb}
\usepackage{graphicx}% Include figure files
\usepackage{caption}
\usepackage{color}% Include colors for document elements
\usepackage{dcolumn}% Align table columns on decimal point
\usepackage{bm}% bold math
\usepackage{float}
\usepackage{hyperref} % hyperref must be loaded before apacite
\usepackage{apacite}
\usepackage{todonotes}
\usepackage{comment}

\usepackage[utf8]{inputenc} 
\usepackage{titling}

\definecolor{background-color}{gray}{0.98}

\newcommand{\q}[1]{\lq\lq{}{}#1\rq\rq{}{}}
\newcommand*\samethanks[1][\value{footnote}]{\footnotemark[#1]}
\thanksmarkseries{alph}

\title{Bias in Data-driven AI Systems -\\ An Introductory Survey}

\author{
    Eirini Ntoutsi\thanks{L3S Research Center \& Leibniz University Hannover, Germany - \textit{ntoutsi@l3s.de}}, 
    Pavlos Fafalios\thanks{Institute of Computer Science, FORTH-ICS, Greece},
    Ujwal Gadiraju\samethanks[1], 
    Vasileios Iosifidis\samethanks[1],\\
    Wolfgang Nejdl\samethanks[1], 
    Maria-Esther Vidal\thanks{TIB Leibniz Information Centre For Science and Tecnhnology, Hannover, Germany},
    Salvatore Ruggieri\thanks{KDDLAB, Dipartimento di Informatica, Università di Pisa, Italy},
    Franco Turini\samethanks,\\
        Symeon Papadopoulos\thanks{Information Technologies Institute, CERTH, Thessaloniki, Greece},
    Emmanouil Krasanakis\samethanks,
    Ioannis Kompatsiaris\samethanks,\\
    Katharina Kinder-Kurlanda\thanks{GESIS Leibniz Institute for the Social Sciences, Cologne, German},
    Claudia Wagner\samethanks,
    Fariba Karimi\samethanks,
    Miriam\\Fernandez\thanks{Knowledge Media Institute, The Open University, Milton Keynes, UK },
    Harith Alani\samethanks,
    Bettina Berendt\thanks{Faculty of Electrical Engineering and Computer Science, TU Berlin, Germany, and Department of Computer Science, KU Leuven, Belgium},
    Tina Kruegel\thanks{Institute for Legal Informatics, Leibniz University of Hanover, Germany},
    Christian Heinze\samethanks,\\
    Klaus Broelemann\thanks{Innovation Lab, SCHUFA Holding AG, Wiesbaden, Germany},
    Gjergji Kasneci\samethanks, 
    Thanassis Tiropanis\thanks{Electronics and Computer Science, University of Southampton, UK},~
    Steffen Staab$^{a,k,}$\thanks{IPVS, Universit\"at Stuttgart, Germany}
}
\date{}
\begin{document}
\maketitle

\begin{center}
\subsubsection*{\small Article Type:}
Overview
%Opinion, Primer, Overview, Advanced Review, Focus Article, or Software Focus
%The Article Type denotes the intended level of readership for your article. An Editor may have mentioned a specific Article Type in your invitation letter; if so, please let them know if you think a different Article Type better suits your topic.

\hfill \break
%\thanks

\subsubsection*{Abstract}
%The abstract should not exceed 250 words and should be a concise description of the article and its implications. It should include all keywords associated with your article, as keywords increase its discoverability. Please do not include phrases such as ``This article discusses \ldots" or ``Here we review \ldots", references to other articles, or URLs.

\begin{flushleft}
AI-based systems are widely employed nowadays to make decisions that have far-reaching impacts on individuals and society. Their decisions might affect everyone, everywhere and anytime,
entailing concerns about potential human rights issues.
Therefore, it is necessary to move beyond traditional AI algorithms optimized for predictive performance and embed ethical and legal principles in their design, training and deployment to ensure social good while still benefiting from the huge potential of the AI technology.
The goal of this survey is to provide a broad multi-disciplinary overview of the area of bias in AI systems, focusing on technical challenges and solutions as well as to suggest  new research directions towards approaches well-grounded in a legal frame.
In this survey, we focus on data-driven AI, as a large part of AI is powered nowadays by (big) data and powerful Machine Learning (ML) algorithms. 
If otherwise not specified, we use the general term bias to describe problems related to the gathering or processing of data that might result in  prejudiced decisions on the bases of demographic features like race, sex, etc.
\end{flushleft}
\end{center}

\section*{\sffamily \Large Introduction}
Artificial Intelligence (AI) algorithms are widely employed %nowadays 
by businesses, governments, and other organisations in order to make decisions that have far-reaching impacts on individuals and society. 
Their decisions might influence everyone, everywhere and anytime, offering solutions to %important
problems faced in different disciplines or in daily life, but at the same time entailing risks like being denied a job or a medical treatment.
The discriminative impact of AI-based decision making to certain population groups has been already observed in a variety of cases.
For instance, the COMPAS system for predicting the risk of re-offending was found to predict higher risk values for black defendants (and lower for white ones) than their actual risk \cite{angwin2016machine} (\textit{racial-bias}). In another case, Google’s Ads tool for targeted advertising was found to serve significantly fewer ads for high paid jobs to women than to men \cite{datta2015automated} (\textit{gender-bias}). 
Such incidents have led to an ever increasing public concern about the impact of AI in our lives. 

\begin{figure}[t]
\includegraphics[width=\textwidth, trim={1cm 6.5cm 1cm 6cm}, clip]{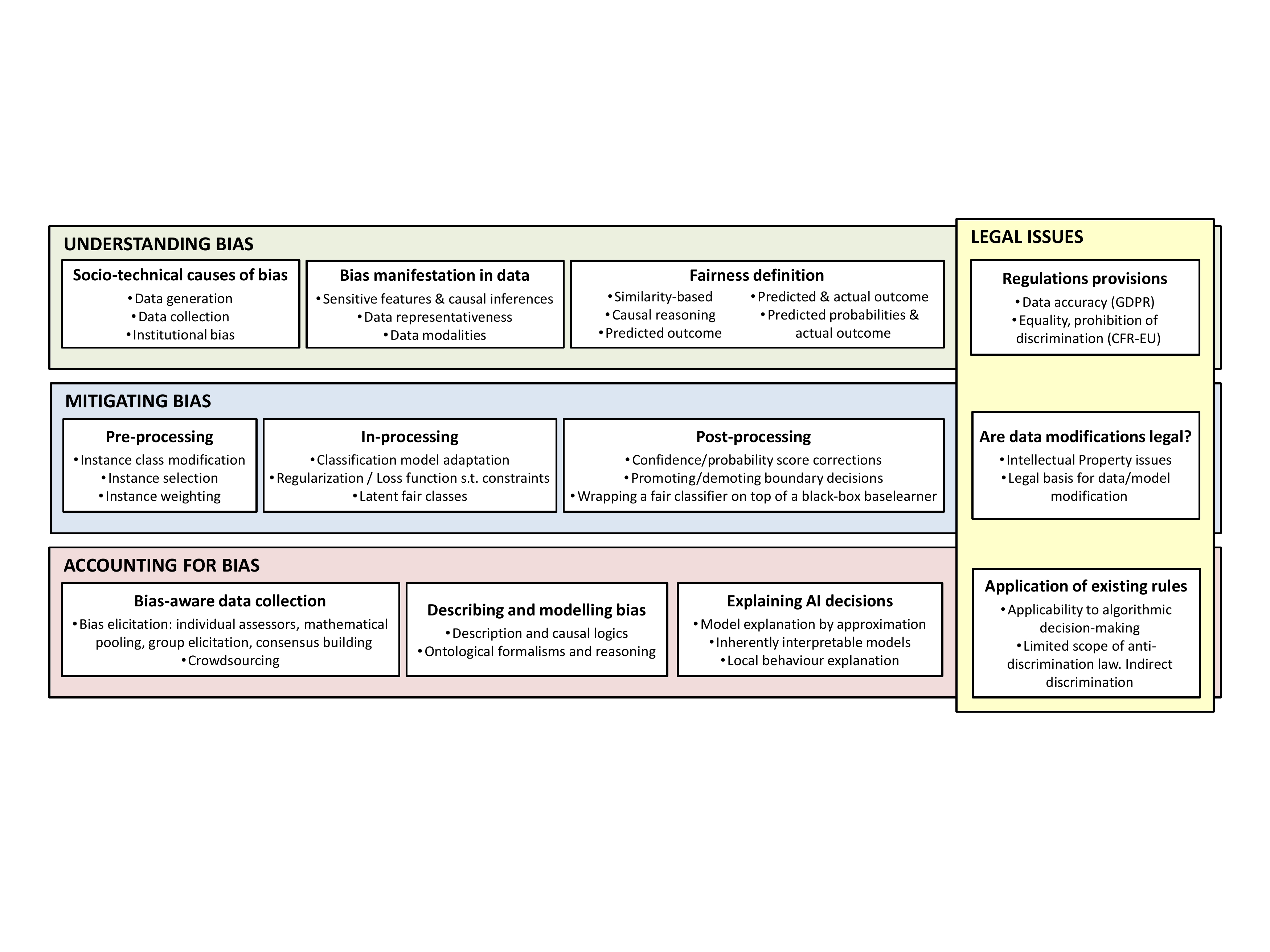}
\caption{Overview of topics discussed in this survey.}
\label{fig:overview}
\end{figure}

Bias is not a new problem rather \textit{\q{bias is as old as human civilization}} and \textit{\q{it is human nature for members of the dominant majority to be oblivious to the experiences of other groups}}.\footnote{Fei-Fei Li, Chief-Scientist for AI at Google and Professor at Stanford (\url{http://fortune.com/longform/ai-bias-problem/}).} 
However, AI-based decision making may magnify pre-existing biases and evolve new classifications and criteria with huge potential for new types of biases. 
These constantly increasing concerns have led to a reconsideration of AI-based  
systems towards new approaches that also address the fairness of their decisions. 
In this paper, we survey recent technical approaches on bias and fairness in AI-based decision making systems, we discuss their legal ground\footnote{The legal discussion in this paper refers primarily to the EU situation. We acknowledge the difficulty of mapping the territory between AI and the law as well as the extra complexity that country-specific legislation brings upon and therefore, we believe this is one of the important areas for future work.} as well as open challenges and directions towards AI-solutions for societal good.
We divide the works into three broad categories:

\begin{itemize}  
    \item \textit{Understanding bias}. Approaches that help understand how bias is created in the society and enters our socio-technical systems, is manifested in the data used by AI algorithms, and can be modelled and formally defined.

    \item \textit{Mitigating bias.}  Approaches that tackle bias in different stages of AI-decision making, namely, pre-processing,
     in-processing and post-processing methods focusing on data inputs, learning algorithms and model outputs, respectively. 
    
    \item  \textit{Accounting for bias.} Approaches that account for bias proactively, via bias-aware data collection,
    or  retroactively, by explaining AI-decisions in human terms.
\end{itemize}
Figure \ref{fig:overview} provides a visual map of the topics discussed in this survey.

This paper complements existing surveys that either have a strong focus on machine ethics, such as \cite{yu2018building}, study a specific sub-problem, such as explaining black box models \cite{DBLP:journals/csur/GuidottiMRTGP19,atzmueller2017declarative}, 
or focus in specific contexts, such as the Web~\cite{baeza2018bias}, by providing a broad categorisation of the technical challenges and solutions, a comprehensive coverage of the different lines of research as well as their legal grounds.

We are aware that the problems of bias and discrimination are not limited to AI and that the technology can be deployed (consciously or unconsciously) in ways that reflect, amplify or distort real world perception and status quo. Therefore, as the roots to these problems are not only technological, it is naive to believe that technological solutions will suffice.
Rather, more than technical solutions are required including socially acceptable definitions of fairness and meaningful interventions to ensure the long-term well-being of all groups. These challenges require multi-disciplinary perspectives and a constant dialogue with the society as bias and fairness are multifaceted and volatile.
Nevertheless, as the AI technology penetrates in our lives, it is extremely important for technology creators to be aware of bias and discrimination and to ensure responsible usage of the technology, keeping in mind that a technological approach on its own is not a panacea for all sort of bias and AI problems.
%========================

\section*{\sffamily \Large Understanding Bias}
\label{sec:understanding}
Bias is an old concept in Machine Learning (ML), traditionally referring to the assumptions made by a specific model (\emph{inductive bias})~\cite{Mitchell:1997:ML:541177}. A classical example is Occam's razor preference for the simplest hypothesis. With respect to human bias, its many facets have been studied by many disciplines including psychology, ethnography, law, etc.
In this survey, we consider as bias the \textit{inclination or prejudice of a decision made by an AI system which is  for or against one person or group, especially in a way considered to be unfair}. Given this definition, we focus on how bias enters AI systems %(Sect.~\ref{sec-bias-society}) 
and how it is manifested in the data comprising the input to AI algorithms. %(Sect.~\ref{sec:bias_in_data}).
Tackling bias entails answering the question how to define fairness such  that it can be  considered in AI systems; we discuss different fairness notions employed by existing solutions.
Finally, we close this section with legal implications of data collection and bias definitions.

\subsection*{\sffamily \large Socio-technical causes of bias}
\label{sec-bias-society}
AI relies heavily on data generated by humans (e.g., user-generated content) 
or collected via systems created by humans. Therefore, whatever biases exist in humans enter our systems and even worse, they are amplified due to the complex sociotechnical systems, such as the Web.\footnote{The Web Science Manifesto - By Web Scientists. Building Web Science. For a Better World, \url{https://www.webscience.org/manifesto/}}  
As a result, algorithms may reproduce (or even increase) existing inequalities or discriminations \cite{karimi2018homophily}. Within societies, certain social groups may be disadvantaged, which usually results in \q{institutional bias} 
where there is a tendency for the procedures and practices of particular institutions to operate in ways in which some social groups are being advantaged and others disadvantaged. 
This need not be the result of conscious discrimination but rather of the majority following existing norms. Institutional racism and sexism are common examples \cite{chandler2011dictionary}. 
Algorithms are part of existing (biased) institutions and structures, but they may also amplify or introduce bias as they favour those phenomena and aspects of human behaviour that are easily quantifiable over those which are hard or even impossible to measure. This problem is exacerbated by the fact that certain data may be easier to access and analyse than others, which has caused, for example, the role of Twitter for various societal phenomena to be overemphasized \cite{tufekci2014big}. Once introduced, algorithmic systems encourage the creation of very specific data collection infrastructures and policies, for example, they may suggest tracking and surveillance \cite{introna2004picturing} which then change or amplify power relations. Algorithms thus shape societal institutions and potential interventions, and vice versa.
It is currently not entirely clear, how this complex interaction between algorithms and structures plays out in our societies. Scholars have thus called for \q{algorithmic accountability} to improve understanding of the power structures, biases, and influences that algorithms exercise in society \cite{diakopoulos2015algorithmic}.

\subsection*{\sffamily \large How is bias manifested in data?}
\label{sec:bias_in_data}
Bias can be manifested in (multi-modal) data through sensitive features and their causal influences, or through under/over-representation of certain groups. 

\subsubsection*{\sffamily \normalsize Sensitive features and causal influences}
Data encode a number of people characteristics in the form of feature values. Sensitive characteristics that identify grounds of discrimination or bias may be present or not. Removing or ignoring such sensitive features does not prevent learning biased models, because other correlated features (also known as \textit{redundant encodings}) may be used as proxies for them. For example, neighbourhoods in US cities  are highly correlated with race, and this fact has been used for systematic denial of services such as bank loans or same-day purchase delivery.\footnote{Amazon doesn’t consider the race of its customers. Should It? \url{https://www.bloomberg.com/graphics/2016-amazon-same-day/}} Rather, including sensitive features in data may be beneficial in the design of fair models \cite{DBLP:journals/ail/ZliobaiteC16}. Sensitive features may also be correlated with the target feature that classification models want to predict. For example,  a minority’s preference for red cars may induce bias against the minority in predicting accident rate if red cars are also preferred by aggressive drivers. Higher insurance premium may then be set for red car owners, which disproportionately impacts on minority members. Simple correlation between apparently neutral features can then lead to biased decisions. 
Discovering and understanding causal influences among variables is a fundamental tool for dealing with bias, as recognised in the legal circles \cite{Foster2004} and in medical research \cite{Grimes2002}. The interested reader is referred to the recent survey on causal approaches to fairness in classification models
\cite{DBLP:journals/corr/abs-1805-05859}.

\subsubsection*{\sffamily \normalsize Representativeness of data}
Statistical (including ML) inferences require that the data from which the model was learned be representative of the data on which it is applied.
However, data collection often suffers from biases that lead to the over- or under-representation of certain groups, 
especially in big data, where many
data sets have not been created with the rigour of a 
statistical study,  
but they are the by-product of other activities with different, often operational, goals \cite{BarocasSelbst16}.  
Frequently occurring biases 
include {\em selection
bias} (certain individuals are more likely to be selected for study),
often as {\em self-selection bias}, and the reverse {\em exclusion bias}; 
{\em reporting bias}
(observations of a certain kind are more likely to be reported, which leads to a sort of selection bias on observations); 
and
{\em detection bias}  
(a phenomenon is more likely to be observed for a particular set of subjects).
Analogous biases can lead to under- or over-representations of properties of individuals, e.g., \cite{boyd2012critical}. 
If the mis-represented groups coincide with social groups against which there already exists social bias such as prejudice or discrimination, even ``unbiased computational processes can lead to discriminative decision procedures'' \cite{DBLP:series/sapere/CaldersZ13}. 
Mis-representation in the data can lead to vicious cycles that perpetuate discrimination and disadvantage \cite{BarocasSelbst16}.
Such ``pernicious feedback loops''
\cite{O'Neil:2016:WMD:3002861}
can occur with both under-representation of historically disadvantaged groups, e.g., women and people of colour in IT developer communities and image datasets \cite{buolamwini2018gender}, and
with over-representation, e.g., black people in drug-related arrests \cite{LumIsaac}. 

\subsubsection*{\sffamily \normalsize Data modalities and bias}
Data comes in different modalities (numerical, textual, images, etc.) as well as in multimodal representations (e.g., audio-visual content).
Most of the fairness-aware ML approaches refer to  structured data represented in some fixed feature space. %(c.f. Sect.~\ref{sec:mitigating}).
Data modality-specific approaches also exist, especially for textual data and images.
Bias in language has attracted a lot of recent interest with many studies exposing a large number of offensive associations related to gender and race on publicly available word embeddings~\cite{bolukbasi2016man} as well as how these associations have evolved over time~\cite{C18-1117}.
Similarly for the computer vision community where standard image collections like MNIST are exploited for training, or off-the-shelf pre-trained models are used as feature extractors, assuming the collections comprise representative samples of the real world. In reality, though, the collections can be biased as many recent studies have indicated.
For instance, \cite{buolamwini2018gender} have found that commercial facial recognition services perform much better on lighter male subjects than darker female ones.
Overall, the additional layer of feature extraction that is typically used within AI-based multimodal analysis systems makes it even more challenging to trace the source of bias in such systems.

\label{fairness_measures}
\subsection*{\sffamily \large How is fairness defined?}
\label{sec:definitions}
More than 20 different definitions of fairness have appeared thus far in the computer science   literature~\cite{DBLP:conf/icse/VermaR18,DBLP:journals/datamine/Zliobaite17}; and some of these definitions and others were proposed and investigated in work on formalizing fairness from other disciplines, such as education, over the past 50 years~\cite{DBLP:conf/fat/HutchinsonM19}. 
Existing fairness definitions can be categorized into: (i) ``predicted outcome", (ii) ``predicted and actual outcome", (iii) ``predicted probabilities and actual outcome", (iv) ``similarity based" and (v) ``causal reasoning"~\cite{DBLP:conf/icse/VermaR18}.  
``Predicted outcome" definitions solely rely on a model's predictions (e.g., \textit{demographic parity} checks the percentage of protected and non-protected groups in the positive class). ``Predicted and actual outcome" combine a model's predictions with the true labels 
(e.g., \textit{equalized odds} requires false positive and negative rates to be similar amongst protected and non-protected groups). ``Predicted probabilities and actual outcome" employ the predicted probabilities instead of the predicted outcomes (e.g., \textit{good calibration} requires the true positive probabilities between protected and non-protected groups to be the same). Contrary to definitions (i)-(iii) that only consider the sensitive attribute, ``similarity based" definitions also employ non-sensitive attributes (e.g., \textit{fairness through awareness} states that similar individuals must be treated equally). 
Finally, ``causal reasoning" definitions are based on directed acyclic graphs that capture relations between features and their impact on the outcomes by structural equations (e.g.,  \textit{counterfactual fairness}~\cite{kusner2017counterfactual} constructs a graph that verifies whether the attributes defining the outcome are correlated to the sensitive attribute).

Despite the many formal, mathematical definitions of fairness proposed over the last years the problem of formalising fairness is still open as well as the discussion about the merits and demerits of the different measures. \cite{corbett2018measure} show the statistical limitations of prevailing mathematical definitions of fairness and the (negative) effect of enforcing such fairness-measures on group well-being and urge the community to explicitly focus on consequences of potential interventions.  

\subsection*{\sffamily \large Legal issues of bias and fairness in AI}
\label{sec:legal_on_bias_in_data}

Taking into account the variety of bias creation in AI systems and its impact on society, the question arises whether the law should provide regulations for non-discriminatory AI- based decision making.
Generally speaking, existing EU regulation comes into play when (discriminatory) decisions have been taken, while provisions tackling the quality of selected data are rare. For the earlier, the control of discriminatory decisions, the principle of equality and the prohibition of discrimination (Art. 20, 21 EU Charter of Fundamental Rights, Art. 4 Directive 2004/113 and other directives) apply. However, these provisions only address discrimination on the basis of specific criteria and require prima facie evidence of a less favourable treatment on grounds of a prohibited criterion, which will often be difficult to establish~\cite{hacker2018teaching}. For the latter, the control of the quality of the selected data, with respect to “personal data” Art. 5 (1) GDPR,\footnote{The General Data Protection Regulation (EU) 2016/679 (GDPR), \url{https://gdpr-info.eu/}} stipulates “the principle of data accuracy” which, however, does not hinder wrongful or disproportionate selection. 
With respect to automated decision-making (Art. 22 GDPR), recital 71 only points out that appropriate mathematical or statistical procedures shall be used and that discriminatory effects shall be prevented. While the effectiveness of Art. 22 GDPR is uncertain~\cite{zuiderveen2018discrimination} 
it provides some safeguards, such as restrictions on the use of automated decision-making, and, where it is used, a right to transparency, to obtain human intervention and to contest the decision. Finally, some provisions in area-specific legislation can be found, e.g., Art. 12 Regulation (EC) No 223/2009 for European statistics.

%========================
\section*{\sffamily \Large Mitigating Bias}
\label{sec:mitigating}
Approaches for bias mitigation can be categorised into: i) pre-processing methods focusing on the data, ii) in-processing methods focusing on the ML algorithm, and iii) post-processing methods focusing on the ML model.
We conclude the section with a discussion on the legal issues of bias mitigation. 

%===========================================
\subsection*{\sffamily \large Pre-processing approaches}
\label{sec:preprocessing}
Approaches in this category focus on the data, the primary source of bias, aiming to produce a ``balanced'' dataset that can then be fed into any learning algorithm. The intuition behind these approaches is that the fairer the training data is, the less discriminative the resulting model will be. 
Such methods modify the original data distribution by altering class labels of carefully selected instances close to the decision boundary~\cite{kamiran2009classifying} or in local neighbourhoods~\cite{DBLP:conf/kdd/ThanhRT11}, 
by assigning different weights to instances based on their group membership~\cite{DBLP:conf/icdm/CaldersKP09} or by carefully sampling from each group.
These methods use heuristics aiming to balance the protected and unprotected groups in the training set; however, their impact is not well controlled despite their efforts for minimal data  interventions.
Recently, \cite{DBLP:conf/nips/CalmonWVRV17} proposed a probabilistic fairness-aware framework that alters the data distribution towards fairness while controlling the per-instance distortion and  by preserving data utility for learning. 

%===========================================
\subsection*{\sffamily \large In-processing approaches}
\label{sec:inprocessing}
In-processing approaches reformulate the classification problem by explicitly incorporating the model's discrimination behaviour in the objective function through regularization or constraints, or by training on latent target labels.
For example, \cite{DBLP:conf/icdm/KamiranCP10} modify the splitting criterion of decision trees to also consider the impact of the split w.r.t. the protected attribute. 
\cite{kamishima2012fairness} integrate a regularizer  to reduce the effect of ``indirect prejudice'' (mutual information between the sensitive features and class labels).
\cite{dwork2012fairness} redefine the classification problem by minimizing an arbitrary loss function subject to the \emph{individual fairness-constraint} (similar individuals are treated similarly). 
\cite{zafar2017fairness} propose a constraint-based approach for \emph{disparate mistreatment} (defined in terms of misclassification rates) which can be incorporated into logistic-regression and SVMs.
In a different direction, \cite{krasanakis2018adaptive} assume the existence of latent fair classes and propose an iterative training approach towards those classes which alters the in-training weights of the instances. 
\cite{IosNto19} propose a sequential fair ensemble, AdaFair, that extends the weighted distribution approach of AdaBoost by also considering the cumulative fairness of the learner up to the current boosting round and moreover, it optimises for balanced error instead of overall error to account for class imbalance.

While most of the in-processing approaches refer to classification, approaches for the unsupervised case have also emerged recently,  for example, the fair-PCA approach of \cite{DBLP:conf/nips/SamadiTMSV18} that forces equal reconstruction errors for both protected and unprotected groups. \cite{chierichetti2017fair} formulate the problem of fair clustering as having  approximately equal representation for each protected group in every cluster and define fair-variants of classical  $k$-means and $k$-medoids algorithms.
%===========================================
\subsection*{\sffamily \large Post-processing approaches}
\label{sec:postprocessing}
The third strategy is to post-process the classification model once it has been learned from data. This consists of altering the model's internals (white-box approaches) or its predictions (black-box approaches). Examples of the white-box approach consist of correcting the confidence of CPAR classification rules \cite{DBLP:conf/sdm/PedreschiRT09}, probabilities in Na\"ive Bayes models \cite{DBLP:journals/datamine/CaldersV10}, or the class label at leaves of decision trees \cite{DBLP:conf/icdm/KamiranCP10}. 
White-box approaches have not been further developed in recent years, being superseded by in-processing methods. 
Examples of the black-box approach aim at keeping proportionality of decisions among protected vs unprotected groups by promoting or demoting predictions close to the decision boundary \cite{DBLP:journals/isci/KamiranMKZ18}, by differentiating the decision boundary itself over groups \cite{DBLP:conf/nips/HardtPNS16},
or by wrapping a fair classifier on top of a black-box base classifier \cite{DBLP:conf/icml/AgarwalBD0W18}. An analysis of how to post-process group-wise calibrated classifiers under fairness constraints is given in \cite{Canetti:2019:SCH:3287560.3287561}.
While the majority of approaches are concerned with classification models, bias post-processing has been deemed as relevant when interpreting clustering models as well \cite{Lorimer2017}.

\subsection*{\sffamily \large Legal issues of mitigating bias}
\label{subsec:legalMitigation} 
% AKIS: The phrasing of this paragraph is not very clear imho. I understand that in general there are no legal issues specific to bias regarding the different mitigation approaches. We essentially say that only IP and data protection regulations apply, but is this relevant to the discussion? Perhaps, there's room for shortening and simplifying this paragraph.

%From a legal point of view it is to be questioned 

Pertinent legal questions involve whether modifications of data as envisaged by the pre-and in-processing approaches, as well as altering the model in the post-processing approach, could be considered lawful. Besides intellectual property issues that might occur, there is no general legal provision dealing with the way data is collected, selected or (even) modified. Provisions are in place mainly if such training data would (still) be personal data. Modifications (as well as any other processing) would need a legal basis. However, legitimation could derive from informed consent (provided that specific safeguards are met), or could rely on contract or legitimate interest. Besides, data quality could be relevant in terms of warranties, if a data provider sells data. A specific issue arises when \lq{}debiasing\rq{} involves sensitive data, as under Art. 9 GDPR special category data such as ethnicity often requires explicit consent~\cite{kilbertus2018blind}. A possible solution could be Art. 9(2)(g) GDPR which permits processing for reasons of substantial public interest, which arguably could be seen in ‘debiasing’. The same grounds of legitimation apply when altering the model. 
However, contrary to data modification, data protection law would arguably not be applicable here, as the model would not contain personal data, unless the model is vulnerable to confidentiality attacks such as model inversion and membership inference~\cite{veale2018algorithms}.

\section*{\sffamily \Large Accounting for Bias}
\label{sec:accounting}
Algorithmic accountability refers to the assignment of responsibility for how an algorithm is created and its impact on society \cite{algoaccountability18}. 
In case of AI algorithms the problem is aggravated as we do not codify the solution, rather the solution is inferred via machine learning algorithms and complex data.
AI accountability has many facets, we focus below on the most prominent ones that account for bias either \textit{proactively}, via bias-aware data collection, or \textit{retroactively} by explaining AI decisions in human terms; furthermore, we discuss the importance of describing and documenting bias by means of formalisms like ontologies.

%===========================================

\subsection*{\sffamily \large Proactively: Bias-aware data collection}
\label{sec:data-collection}
A variety of methods are adopted for data acquisition to serve diverse needs; these may be prone to introducing bias at the data collection stage itself, e.g., \cite{morstatter2014biased}. Proposals have been made for a structured approach to bias elicitation in evidence synthesis, including bias checklists and elicitation tasks that can be performed either by individual assessors and mathematical pooling, group elicitation and consensus building or hybrid approaches \cite{turner:08}. However, bias elicitations have themselves been found to be biased even when high quality assessors are involved and remedies have been proposed \cite{manzi:18}. 

Among other methods, crowdsourcing is a popular approach that relies on large-scale acquisition of human input for dealing with data and label scarcity in ML. 
Crowdsourced data and labels may be subject to bias at different stages of the process: 
task design and experimental setup, task decomposition and result aggregation, selection of workers and the entailing human factors  ~\cite{hube2019understanding,DBLP:conf/nips/KargerOS11,kamar2015identifying}.
Mitigating biases in crowdsourced data becomes harder in subjective tasks, where the presence of varying ideological and cultural backgrounds of workers means that it is possible to observe biased labels with complete agreement among the workers.

\subsection*{\sffamily \large Describing and modeling bias using ontologies}
\label{sec:describingBias}
Accounting for bias not only requires  understanding of the different sources, i.e., data, knowledge bases, and algorithms, but more importantly, it demands the interpretation and description of the meaning, potential side effects, provenance, and context of bias. 
Usually unbalanced categories are understood as bias and considered as sources of negative side effects. Nevertheless, skewed distributions may simply hide features or domain characteristics that, if removed, would hinder the discovery of relevant insights. This situation can be observed, for instance, in populations of lung cancer patients. As highlighted in diverse scientific reports, e.g., \cite{Garrido2018}, lung cancer in women and men has significant differences such as aetiology, pathophysiology, histology, and risk factors, which may impact in cancer occurrence, treatment outcomes, and survival. Furthermore, there are specific organizations that collaborate in lung cancer prevention and in the battle against smoking; some of these campaigns are  oriented to particular focus groups and the effects of these initiatives are observed in certain populations. All these facts impact on the gender distribution of the population and could be interpreted as bias. However, in this context, imbalance reveals domain specific facts that need to be preserved in the population, and a formal description of these uneven distributions should be provided to avoid misinterpretation. Moreover, as any type of data source, knowledge bases and ontologies can also suffer from various types of bias or  knowledge imbalance. For example, the description of the existing mutations of a gene in a knowledge base like COSMIC,\footnote{\url{https://cancer.sanger.ac.uk/cosmic}} or the properties associated with a gene in the Gene Ontology (GO),\footnote{\url{http://geneontology.org/}} may be biased by the amount of research that has been conducted in the diseases associated with these genes. Expressive formal models are demanded in order to describe and explain the characteristics of a data source and under which conditions or context, the data source is biased. 

Formalisms like description and causal logics, e.g., \cite{BesnardCM14,DehaspeR96,Krotzsch0OT18,LeBlancBV19}, allow for measuring and detecting bias in data collections of diverse types, e.g, online data sets \cite{PitouraTFFPAW17} and recommendation systems \cite{SerbosQMPT17}. They also enable the annotation of statements with trustworthiness \cite{SonPB15} and temporality \cite{OzakiKR19}, as well as causation relationships between them \cite{LeBlancBV19}. Ontologies also play a relevant role as knowledge representation models for describing universe of discourses in terms of concepts such as classes, properties, and subsumption relationships, as well as contextual statements of these concepts. 
NdFluents~\cite{Gimenez-GarciaZ17} and Context Ontology Language (CoOL) \cite{StrangLF03}, represent exemplar ontology formal models able to  express and combine diverse contextual dimensions and interrelations (e.g., locality and vicinity). Albeit expressive, existing logic-based and ontological formalisms are not tailored for representing contextual bias or differentiating unbalanced categories that consistently correspond to instances of a real-world domain. Therefore, expressive ontological formalisms are demanded to represent the contextual dimensions of various types of sources, e.g., data collections, knowledge bases, or ontologies, as well as annotations denoting causality and provenance of the represented knowledge. These formalisms will equip bias detection algorithms with reasoning mechanisms that not only enhance accuracy but also enable explainability of the meaning, conditions, origin, and context of bias. Thus, domain modelling using ontologies will support context-aware bias description and interpretability.

\subsection*{\sffamily \large Retroactively: Explaining AI decisions}
\label{sec:explainability}
% AKIS: This section has far too many references (all very relevant and interesting, but currently unbalanced compared to other sections).
% maybe replace some of those with more review-style papers like: https://arxiv.org/abs/1708.08296 ?

%\begin{itemize}
%    \item of white models [SCHUFA](Person(-s)?)
%    \item of black models [UNIPI] (Person(-s)?)
%\end{itemize} 
Explainability refers to the extent the internal mechanics of a learning model can be explained in human terms. It is often used interchangeably with interpretability, although the latter refers to whether one can predict what will happen given a change in the model input or parameters.
% This part contributed by UNIPI (Franco and Salvatore)
Although attempts to tackle  \emph{interpretable} ML have existed for some time %several years now 
\cite{DBLP:journals/expert/HoffmanK17}, %ercim2019, ,DBLP:conf/dsaa/GilpinBYBSK18 DBLP:journals/corr/abs-1901-04592
there has been an exceptional growth of research literature in the last years with  emerging keywords such as \emph{explainable AI} \cite{DBLP:journals/access/AdadiB18} and \emph{black box explanation}~\cite{DBLP:journals/csur/GuidottiMRTGP19}. 
Many papers propose approaches for understanding the \textit{global} logic of a model by building an interpretable classifier able to mimic the obscure decision system.
Generally, these methods are designed for explaining specific models, e.g.,~deep neural networks~\cite{DBLP:journals/dsp/MontavonSM18}.
Only few are agnostic to the black box model~\cite{DBLP:journals/datamine/HeneliusPBAP14}.
The difficulties in explaining black boxes and complex models \textit{ex-post}, have  motivated proposals of transparent classifiers which are interpretable on their own and exhibit predictive accuracy close to that of obscure models. These include Bayesian models \cite{DBLP:conf/kdd/0013H17}, generalized additive models \cite{DBLP:conf/kdd/LouCGH13}, supersparse linear models \cite{DBLP:journals/ml/UstunR16},  rule-based decision sets \cite{DBLP:conf/kdd/LakkarajuBL16}, optimal classification trees \cite{DBLP:journals/ml/BertsimasD17}, model trees \cite{broelemann2019modeltrees} and neural networks with interpretable layers~\cite{zhang2018interpretable}.

A different stream of approaches focuses on the \textit{local} behavior of a model, searching for an explanation of the decision made for a specific instance \cite{DBLP:journals/csur/GuidottiMRTGP19}. Such approaches are either \textit{model-dependent}, e.g., Taylor approximations~\cite{Kasneci2016}, saliency masks (the image regions that are mainly responsible for the decision) for neural network decisions ~\cite{DBLP:conf/cyberc/MaYY15}, and attention models for recurrent networks~\cite{choi2016retain}, 
or \textit{model-agnostic}, such as those started by the LIME method~\cite{DBLP:conf/kdd/Ribeiro0G16}. 
% AKIS: are both Ribeiro's citations necessary here?
% Salvatore: removed this one ,DBLP:conf/aaai/Ribeiro0G18
The main idea is to derive a local explanation for a decision outcome on a specific instance by learning an interpretable model from a randomly generated neighbourhood of the instance.
A third stream aims at bridging the local and the global ones by defining a strategy for combining local models in an incremental way \cite{pedreschi2019}.
%
% \todo{\textbf{Steffen}: I find the following paragraph hard to understand. It does not fit neatly into the narrative and I do not know what it adds beyond listing the citations.}
% Since explanations must be comprehensible to humans
% (possibly with different levels of expertise), a key issue regards the degree of interpretability of a model \cite{DBLP:journals/corr/abs-1811-11839}. % \cite{DBLP:journals/corr/abs-1802-00682}
% User surveys and human-in-the-loop approaches \cite{DBLP:conf/nips/LageRGKD18} suffer from
% a scarcity of observations. Ways that are more practical include measures of model complexity (nodes/depth in a tree, number of rules, statistical information criteria, etc.) \cite{DBLP:phd/de/Ruping2006}.
%To sum up, despite the soaring attention to the topic, the state of the art to date still exhibits ad-hoc, scattered results. A widely applicable, systematic approach with a real impact has not emerged yet. 
% End of contribution by UNIPI (Franco and Salvatore)
%
% Contribution by Southampton (Steffen)
More recent work has  asked the fundamental question \emph{What is an explanation?} \cite{Mittelstadt:2019:EEA:3287560.3287574} and reject such usage of the term `explanation', criticizing that it might be appropriate for a modelling expert, but not for a lay man, and that, for example, humanities or philosophy have an entirely different understanding of what explanations are.

We speculate that there are computational methods that will allow us to find some middle ground. For instance, 
some approaches in ML, 
statistical relational learning in particular \cite{raedt2016statistical},  take the perspective of knowledge representation and reasoning into account  when developing ML models on more formal logical and statistical
grounds. 
AI knowledge representation has been developing a rich theory of argumentation over the last 25 years \cite{DBLP:journals/ai/Dung95}, which recent approaches \cite{DBLP:conf/comma/CocarascuT16} 
try to leverage for generalizing the reasoning aspect of 
 ML towards
the use of computational models of argumentation. The outcome are models of arguments and counterarguments towards certain classifications that can be inspected by a human user
and might be used as formal grounds for explanations 
in the manner that \cite{Mittelstadt:2019:EEA:3287560.3287574} called out for.

\subsection*{\sffamily \large Legal issues of accounting for bias}
While data protection rules affect both the input (data) and the output (automated decision) level of AI decision-making, anti-discrimination laws, as well as consumer and competition rules, address discriminatory policies primarily from the perspective of the (automated) decision and the actions based on it. However, the application of these rules to AI-based decisions is largely unclear. Under present law and the principle of private autonomy, decisions by private parties normally do not have to include reasons or explanations. Therefore, a first issue will be how existing rules can be applied to algorithmic decision-making. Given that a decision will often not be reasoned (hence the reasons will be unknown), it will be difficult to establish that it was made on the basis of a biased decision-making process \cite{Mittelstadt:2019:EEA:3287560.3287574}. 

Even if bias can be proven, a second issue is the limited scope of anti-discrimination law. Under present law, only certain transactions between private parties fall under the EU anti-discrimination directives \cite{LiddellOFlaherty2018}. Moreover, in most cases AI decision-making instruments will not directly use an unlawful criterion (e.g., gender) as a basis for their decision, but rather a ``neutral’’ one (e.g., residence) which in practice lead to a less favourable treatment of certain groups. This raises the difficult concept of indirect discrimination, i.e., a scenario where an ``apparently neutral rule disadvantages a person or a group sharing the same characteristics’’ \cite{LiddellOFlaherty2018}. Finally, most forms of differential treatment can be justified where it pursues a legitimate aim and where the means to pursue that aim are appropriate and necessary. It is unclear whether the argument that AI-based decision making systems produce decisions which are economically sound can be sufficient as justification.

%===========================================
\section*{\sffamily \Large Future directions and conclusions}
There are several directions that can impact this field going forward. 
First, despite the large number of methods for mitigating bias, there are still no conclusive results regarding what is the state of the art method for each category, 
which of the fairness-related interventions perform best, or whether category-specific interventions perform better comparing to holistic approaches that tackle bias at all stages of the analysis process. 
We believe that a systematic evaluation of the existing approaches is necessary to understand their capabilities and limitations and also, a vital part of proposing new solutions. 
The difficulty of the evaluation lies on the fact that different methods work with different fairness notions and are applicable to different AI models.
To this end, benchmark datasets should be made available that cover different application areas and manifest real-world challenges. 
Finally, standard evaluation procedures and measures covering both model performance and fairness-related aspects should be followed, in accordance to international standards like the IEEE - ALGB-WG - Algorithmic Bias Working Group\footnote{https://standards.ieee.org/project/7003.html}.

Second, we recognize that 
``fairness cannot be reduced to a simple self-contained mathematical definition'', ``fairness is dynamic and social and not a statistical issue''.\footnote{\url{https://www.wired.com/story/ideas-joi-ito-insurance-algorithms/}} 
Also, ``fair is not fair everywhere'' \cite{schafer2015fair} meaning that the notion of fairness varies across countries, cultures and application domains. Therefore, it is important to have realistic and applicable fairness definitions for different contexts as well as domain-specific datasets for  method development and evaluation. Moreover, it is important to move beyond the typical training-test evaluation setup and to consider the consequences of potential fairness-related interventions to ensure long-term wellbeing of different groups.
Finally, given the temporal changes of fairness perception, the question of whether one can train models on historical data and use them for current fairness-related problems becomes increasingly pressing.

Third, the related work thus far focuses mainly on supervised learning. In many cases however, direct feedback on the data (i.e., as labels) is not available. Therefore alternative learning tasks should be considered, like unsupervised learning or Reinforcement Learning (RL) where only intermediate feedback is provided to the model. Recent works have emerged in this direction, e.g.,  \cite{jabbari2017fairness} examine fairness in the RL context where one needs to reconsider the effects of short-term actions on long-term rewards.

% AKIS: ML or AI?
Fourth, there is a general trend in the ML community recently for generating plausible data from existing data using Generative Adversarial Networks (GAN) in an attempt to cover the high data demand of modern methods, especially DNNs. Recently, such approaches have been used also in the context of fairness~\cite{xu2018fairgan}, i.e., how to generate synthetic fair data that are similar to the real data. Still however, the problem of representativeness of the training data and its impact on the representativeness of the generated data might aggravate issues of fairness and discrimination.
In the same topic, recent work revealed that DNNs are vulnerable to adversarial attacks, i.e., intentional perturbations of the input examples, and therefore there is a need for methods to enhance their resilience \cite{song2018mat}.

Fifth, AI scientists and everyone involved in the decision making process should be aware of bias-related issues and the effect of their design choices and assumptions.
For instance, studies show that representation-related biases creep into development processes because the development teams are not aware of the importance of distinguishing between certain categories \cite{buolamwini2018gender}. 
Members of a privileged group may not even be aware of the existence of (e.g., racial) categories in the sense that they often perceive themselves as ``just people'', and the interpretation of this as an unconscious default requires the voice of individuals from underprivileged groups, who persistently perceive their being ``different''. % and categorised. 
Two strategies appear promising for addressing this cognitive bias: try to improve diversity in development teams, and subject algorithms to outside and as-open-as-possible scrutiny, for example by permitting certain forms of reverse engineering for algorithmic accountability. %\cite{Diakopoulos13}. 

Finally, from a legal point of view, apart from data protection law, general provisions with respect to data quality or selection are still missing. Recently an ISO standard on data quality (ISO 8000) was published, though not binding and not with regard to decision-making techniques.
Moreover, first important steps have been made, e.g., the Draft Ethics Guidelines for trustworthy AI from the European Commission’s high-level Expert group on AI %\cite{EUDraftEthicsGuidelinesAI} 
or the European parliament resolution containing recommendations to the Commission on Civil Law Rules on Robotics.  %\cite{EUParliamentRoboticsRecommendations}. 
However, these resolutions are still generic. Further interdisciplinary research is needed to define specifically what is needed to meet the balance between the fundamental rights and freedoms of citizens by mitigating bias, while at the same time considering the technical challenges and economical needs. Therefore, any legislative procedures will require a close collaboration of legal and technical experts.
As already mentioned, the legal discussion in this paper refers to the EU where despite the many recent efforts, there is still no consensus for algorithmic fairness regulations across its countries.  Therefore, there is still a lot of work to be done on analysing the legal standards and regulations at a national and international level to support globally legal AI designs.

To conclude, the problem of bias and discrimination in AI-based decision making systems has attracted a lot of attention recently from science, industry, society and policy makers, and there is an ongoing debate on the AI opportunities and risks for our lives and our civilization. This paper surveys technical challenges and solutions as well as their legal grounds in order to advance this field in a direction that exploits the tremendous power of AI for solving real world problems but also considers the societal implications of these solutions.
%\bibliographystyle{named}
%
% SR: named style replaced by these 2 lines: small size + authors' initials instead of full first name
%\clearpage
%
%
As a final note, we want to stress again that biases are deeply embedded in our societies and it is an illusion to believe that the AI and bias problem will be eliminated only with technical solutions.
Nevertheless, as the technology reflects and projects our biases into the future, it is a key responsibility of technology creators to understand its limits and to propose safeguards to avoid pitfalls.
Of equal importance is also for the technology creators to realise that technical solutions without any social and legal ground cannot thrive and therefore multidisciplinary approaches are required.

\section*{Acknowledgement}
This  work  is  supported  by  the  project \textit{\lq\lq{}{}NoBias - Artificial   Intelligence   without   Bias\rq\rq{}{}}, which  has  received  funding from the 
European Union's Horizon 2020 research and innovation programme, under the Marie Skłodowska-Curie (Innovative Training Network) grant agreement  no. 860630.

\bibliography{_BIB.bib}

\end{document}